# Double Compact Binary Merger Rate Density in Open Star Clusters: Black Holes, Neutron Stars, and White Dwarfs

Savannah Cary[1,2], Michiko Fujii[2], Long Wang[3,4], and Ataru Tanikawa[5]
[1] Department of Astronomy, University of California, Berkeley, CA 94720-3411, USA; scary@berkeley.edu
[2] Department of Astronomy, Graduate School of Science, The University of Tokyo, 7-3-1 Hongo, Bunkyo-ku, Tokyo 113-0033, Japan
[3] School of Physics and Astronomy, Sun Yat-sen University, Daxue Road, Zhuhai, 519082, People's Republic of China
[4] CSST Science Center for the Guangdong-Hong Kong-Macau Greater Bay Area, Zhuhai, 519082, People's Republic of China
[5] Center for Information Science, Fukui Prefectural University, 4-1-1 Kenjojima, Matsuoka, Eiheiji-Town, Fukui, 910-1195, Japan

## Abstract

Studying compact object binary mergers in star clusters is crucial for understanding stellar evolution and dynamical interactions in galaxies. Open clusters in particular are more abundant over cosmic time than globular clusters. However, previous research on low-mass clusters with $\lesssim 10^3\,M_\odot$ has focused on binary black holes (BBHs) or black hole–neutron star (BH–NS) binaries. Binary mergers of other compact objects, such as white dwarfs (WDs), are also crucial as progenitors of transient phenomena such as Type Ia supernovae and fast radio bursts (FRBs). We present simulations of three types of open clusters with masses of $10^2$, $10^3$, and $10^4\,M_\odot$. In massive clusters with $\gtrsim 10^4\,M_\odot$, BBHs are dynamically formed; however, less massive compact binaries such as WD–WDs and WD–NSs are perturbed inside the star clusters, causing them to evolve into other objects. We further find BH–NS mergers only in $10^3\,M_\odot$ clusters. Considering star clusters with a typical open cluster mass, we observe that WD–WD merger rates slightly increase for $10^3\,M_\odot$ clusters but decrease for $10^2\,M_\odot$ clusters. Since the host clusters are tidally disrupted, most of them merge outside of the clusters. Our WD–WD merger results have further implications for two classes of transients. Super-Chandrasekhar WD–WD mergers are present in our simulations, demonstrating potential sources of FRBs at a rate of 70–780 $\text{Gpc}^{-3}\,\text{yr}^{-1}$, higher than the rate estimated for globular clusters. Additionally, we find that carbon–oxygen WD–WD mergers in our open clusters (34–640 $\text{Gpc}^{-3}\,\text{yr}^{-1}$) only account for 0.14%–2.6% of the observed Type Ia supernova rate in our local Universe.

*Unified Astronomy Thesaurus concepts:* Stellar mergers (2157); Compact objects (288); Open star clusters (1160)

## 1. Introduction

Compact binary mergers have been of specific interest in recent years. For example, several gravitational-wave signals from neutron star–neutron star (NS–NS), black hole–black hole (BH–BH), and black hole–neutron star (BH–NS) binary mergers have been detected by the LIGO-Virgo-KAGRA collaboration (e.g., R. Abbott et al. 2023a). Furthermore, NS–NS and BH–NS mergers can be used to study short gamma-ray bursts, as they remain the leading progenitors for such explosions (e.g., R. Narayan et al. 1992; E. Berger 2014). Meanwhile, white dwarf–white dwarf (WD–WD) merger events could result in magnetars (e.g., A. R. King et al. 2001; K. Kashiyama et al. 2013), one of the leading theories for the creation mechanism of at least some fast radio bursts (FRBs; Y.-P. Yang & B. Zhang 2018; CHIME/FRB Collaboration et al. 2020; B. Zhang 2020), while carbon–oxygen WD–WD mergers also are believed to be the source of Type Ia supernovae (K. Nomoto 1982; I. J. Iben & A. V. Tutukov 1984; R. F. Webbink 1984; S. Rosswog et al. 2009; S. Cheng et al. 2020). Furthermore, WD–WD mergers can be a source of gravitational radiation (B. Willems et al. 2007; A. J. Ruiter et al. 2010), and are expected to dominate the events detected by future detectors such as the Laser Interferometer Space Antenna (LISA; A. Lamberts et al. 2018; P. Amaro-Seoane et al. 2023). WD–WD mergers may further be a pathway to form pulsars in old stellar systems (K. Kremer et al. 2023, 2024) such as PSR J1922+37, which might be associated with the old open cluster NGC 6791 (X.-J. Liu et al. 2025). Lastly, WD–NS mergers have been considered for transient events such as, but not limited to, FRBs (e.g., X. Liu 2018; S.-Q. Zhong & Z.-G. Dai 2020), gamma-ray bursts (e.g., A. King et al. 2007; J. Yang et al. 2022), and subluminous supernovae (B. D. Metzger 2012; R. Fernández & B. D. Metzger 2013).

To fully understand compact binary merger rates in galaxies, it is important to account for mergers within open clusters. Star clusters contain hundreds to millions of stars, making them an ideal place to study stellar evolution and dynamical interactions, such as these compact binary merger events. Past studies of compact binary mergers typically have focused on the evolution of globular clusters. That being said, there are approximately 150 known globular clusters in the Milky Way, and it is estimated that there are at least 23,000–37,000 open clusters in the Milky Way (S. F. Portegies Zwart et al. 2010). Furthermore, globular clusters were formed 12 to 13 Gyr ago, consisting of old stellar populations. Although open clusters have shorter lifetimes, they can be formed at any age of the Universe. Due to their abundance over cosmic time, open clusters are likely to contribute to the overall compact merger rate in galaxies. Indeed, there have been previous simulation works for BH–BH (e.g., S. Banerjee 2017; U. N. Di Carlo et al. 2020; J. Kumamoto et al. 2020), NS–NS (e.g., G. Fragione & S. Banerjee 2020), and NS–BH (e.g., G. Fragione & S. Banerjee 2020; S. Rastello et al. 2020) merger events in open clusters. There have also been a few simulations of globular and open clusters to study WD–WD mergers and their implications on FRB and Type Ia supernovae production (e.g., K. Kremer et al. 2021a, 2021b, 2023; M. M. Shara & J. R. Hurley 2002). That being said, more work needs to be done for the impact of open clusters on WD–WD and WD–NS systems; our work will comment on such systems in open clusters.

In this paper, we report on compact binary mergers found in N-body simulations of small to large open clusters, while also taking a special interest in high-mass WD–WD mergers. Specifically, we test clusters of size $10^2 M_\odot$, $10^3 M_\odot$, and $10^4 M_\odot$. In Section 2, we describe our simulation methods, and in Section 3 we show our results. These report on the formation efficiencies, rates, and general properties for NS–NS, WD–NS, WD–WD, BH–BH, and BH–NS mergers in open clusters. Furthermore, we use our WD–WD mergers to comment on Type Ia and FRB rates in the local Universe. In Section 4, we conclude our results.

## 2. Methods

We follow the dynamical evolution of open clusters via the N-body simulation code PETAR (L. Wang et al. 2020a). L. Wang et al. (2020a) demonstrate PETAR utilizing the particle–particle particle–tree method (S. Oshino et al. 2011; M. Iwasawa et al. 2017), highly optimized with Framework for Developing Particle Simulators (M. Iwasawa et al. 2016, 2020). Using this method, the force calculation operations in PETAR, $O[N \log N]$, are reduced by a factor of $N$ compared to that of the direct N-body method, $O[N^2]$. PETAR also uses slow-down algorithmic regularization (or SDAR; L. Wang et al. 2020b) method to efficiently and accurately evolve the orbits of close interacting systems, including binaries, triples, and hyperbolic encounters. This code is known to handle the long-term simulations of massive globular clusters with a large fraction of binaries (e.g., J. Wang et al. 2021), as well as open clusters (A. Tanikawa et al. 2024). In PETAR, we also utilize the single and binary stellar evolution package SSE/BSE (J. R. Hurley et al. 2002; S. Banerjee et al. 2020), and the galactic potential package galpy (J. Bovy 2015).

Single stars are evolved according to J. R. Hurley et al. (2000). Massive stars are modeled to collapse to either a BH or NS, and their masses follow the rapid model (C. L. Fryer et al. 2012) with modification of pair-instability supernovae modeled by K. Belczynski et al. (2016). Natal kicks of NSs and BHs are isotropic and given by a Maxwellian distribution centered around 265 km s$^{-1}$ (G. Hobbs et al. 2005). If fallback mass is present, the kick velocity is reduced by $(1 - f_b)$, where $f_b$ is the fraction of the fallback mass (as demonstrated by S. Banerjee et al. 2020). For NSs, if a NS forms via an electron-capture supernova, the kick velocity becomes lower (dispersion ~3 km s$^{-1}$; P. Podsiadlowski et al. 2004). Stellar evolution includes stellar winds formulated by K. Belczynski et al. (2010). Meanwhile, binary evolution follows the model in S. Banerjee et al. (2020) and contains binary evolution processes such as tidal interaction, wind accretion, stable mass transfer, and common envelope evolution.

We simulate open clusters of masses $10^2$, $10^3$, and $10^4 M_\odot$; we generate a total 596, 729, and 17 separate clusters for these initial masses, respectively. The initial conditions are generated using MCLUSTER (A. H. W. Küpper et al. 2011; L. Wang et al. 2019) and are as follows. Each cluster has a starting metallicity of $Z = 0.02$ and an initial binary fraction of 100%. We adopt the Plummer model for the initial phase-space distribution of stars (H. C. Plummer 1911), with initial half-mass radii set to 0.1, 0.4, and 1 pc for masses of $10^2$, $10^3$, and $10^4 M_\odot$, respectively. The initial mass and half-mass–radius of the $10^3 M_\odot$ and $10^4 M_\odot$ models are similar to those of open and young massive clusters in the Milky Way (S. F. Portegies Zwart et al. 2010). For the $10^2 M_\odot$ model, the mass and half-mass–radius follow the typical size of embedded clusters (C. J. Lada & E. A. Lada 2003). For clusters of $10^2$ and $10^3 M_\odot$, the Plummer density distribution is also fractalized, motivated by the fractal features of observed open clusters (S. P. Goodwin & A. P. Whitworth 2004; U. N. Di Carlo et al. 2019). We adopted a fractal dimension of 1.6 following previous studies (U. N. Di Carlo et al. 2019).

The initial mass function (IMF) follows P. Kroupa (2001), with a minimum mass of $0.08 M_\odot$ for all models and a maximum mass of $150 M_\odot$ for the $10^2$ and $10^3 M_\odot$ models and $100 M_\odot$ for the $10^4 M_\odot$ model. The differences in the high-mass limit should not affect our simulations due to the low fraction of stars that would be this massive. For binary systems, the initial distribution of mass ratios, orbital periods, and eccentricities follow H. Sana et al. (2012) for primary masses $>5 M_\odot$ and P. Kroupa (1995a, 1995b) for primary masses $\leqslant 5 M_\odot$. The initial distributions of orbital periods, masses, secondary and primary mass ratios, and eccentricities are shown in Figure 1. The initial eccentricity is truncated at around $10^{-4}$ for the $10^2$ and $10^3 M_\odot$ models, but at $10^{-2}$ for the $10^4 M_\odot$ model. However, the fraction of such low-eccentricity populations is only a few percent of the entire distribution.

Lastly, all clusters are set in a Milky Way potential, with a distance of 8 kpc from the galactic center and a circular velocity of 220 km s$^{-1}$. We follow the evolution of the $10^2 M_\odot$ clusters over the course of 500 Myr, and 1000 Myr for masses of $10^3$ and $10^4 M_\odot$. This is due to the small clusters being completely disrupted by 500 Myr (see Section 3 and Appendix A for more details). Cluster parameters are summarized in Table 1.

We also run isolated binary stellar evolution for each of the 596, 729, and 17 stellar populations generated for our $10^2$, $10^3$, and $10^4 M_\odot$ models, respectively, with the same conditions as primordial binaries in clusters described above. This allows us to observe how isolated binaries evolve compared to the exact same binary population affected by cluster dynamics. We refer to these runs as "BSE."

## 3. Results

### 3.1. Final Status of Cluster

Figure 2 serves as an example of a cluster's appearance at the beginning and end of its simulation for each of the mass conditions. These examples are representative of the typical cluster formed from our initial conditions. Through monitoring the number of objects within the tidal radii of our clusters (refer to Appendix A), we find that by 500 Myr, all $10^2 M_\odot$ clusters are completely dissolved. Therefore, we stop the simulations at 500 Myr. Our $10^3 M_\odot$ clusters are dissolved on timescales $\leqslant 1000$ Myr, where the specific timing is highly dependent on its initial shape due to fractalization (see Figure 6 in Appendix A). Meanwhile, at 1000 Myr, the $10^4 M_\odot$ clusters have not yet dissolved but have lost approximately one-third of stars due to tidal stripping. More information on the full tidal radius evolution for these clusters can be found in Appendix A.

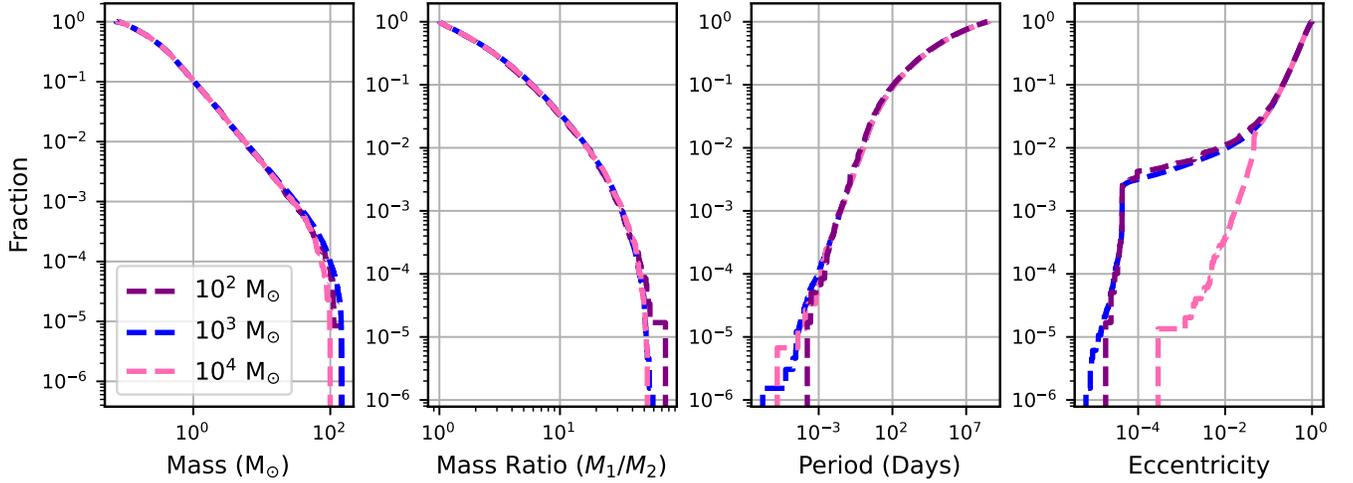

**Figure 1.** Initial properties of stars in simulated clusters. From left to right, we plot the IMFs for each of our three cluster types (combining the stars over all runs into a single IMF per cluster type), and then the initial mass ratios, periods, and eccentricities of our binaries. Note that these distributions are of all stars of all clusters. Although $10^2 \, M_\odot$ clusters appear to have a complete mass function, on a cluster by cluster basis this may not be true.

**Table 1**
Parameters for Our Cluster Models

| Name | $M$ ($M_\odot$) | $r_h$ (pc) | $N_{run}$ | $t_{end}$ (Gyr) | Fractal Dimension | $t_{rh}$ (Myr) | $\rho_c$ ($M_\odot \, pc^{-3}$) | Total Sample Mass ($M_\odot$) |
|---|---|---|---|---|---|---|---|---|
| $10^2 \, M_\odot$ | $10^2$ | 0.1 | 596 | 0.5 | 1.6 | $0.20^{+0.03}_{-0.05}$ | ... | $6.0 \times 10^4$ |
| $10^3 \, M_\odot$ | $10^3$ | 0.4 | 729 | 1 | 1.6 | $3.4^{+0.28}_{-0.33}$ | ... | $7.3 \times 10^5$ |
| $10^4 \, M_\odot$ | $10^4$ | 1.0 | 17 | 1 | Spherical | $33.17^{+0.61}_{-0.81}$ | $3.6 \times 10^3$ | $1.7 \times 10^5$ |

**Note.** We state the initial cluster mass and half-mass–radius for each model. From left to right, we then state the number of simulation runs per cluster model, simulation time limit, fractal dimension, relaxation time, initial core density, and total sample mass ($M \times N_{run}$). Core densities are not calculated for the $10^2$ and $10^3 \, M_\odot$ clusters as they are not uniformly distributed due to fractalization.

### 3.2. Population Properties of Mergers

As described by Figure 3, we look at the merger time, distance, mass, and eccentricity distributions for each type of event. The merger time is defined as the elapsed time between the start of the simulation and the time of the merger event. Eccentricity is calculated by PETAR moments before the merger event. Additionally, a summary of the merger efficiencies (numbers per solar mass) can be found in Table 2.

We look at all mergers produced in the simulation, including those located beyond the cluster tidal radius. All mergers until the end of the simulation are included in our commentary, despite some clusters having been completely dissolved by then. Even if mergers are ejected or the cluster is dissolved, the original cluster may have long-term effects on binary systems.

As we comment on individual merger types, it is important to acknowledge that because of the different initial distributions as highlighted by Figure 1, the BSE may not agree for different cluster masses. For this reason, we keep in mind how BSE and cluster dynamics differ per mass, and use these results to compare the different cluster masses.

In the following, we describe the results of each binary population. However, in our simulations we do not observe a large enough number of NS–NS mergers. Therefore, we do not discuss the distribution of NS–NS mergers.

#### 3.2.1. WD–WD Mergers

The merger efficiency for the $10^2 \, M_\odot$ BSE case is about half of the $10^3$ and $10^4 \, M_\odot$ BSE cases, but this is due to the difference in the simulation time. On the other hand, the difference in merger efficiencies between the $10^3$ and $10^4 \, M_\odot$ BSE models can be understood by the small difference in the initial eccentricity distribution of binaries.

For our cluster simulations, dynamically formed (nonprimordial) mergers make up only about 2%~6% of the total WD–WD mergers between all the cluster masses. For the $10^3 \, M_\odot$ cluster model, ~70% of WD–WD mergers occurred outside of the cluster (>10 pc), while for the $10^4 \, M_\odot$ cluster model, almost 90% of the mergers occurred inside the cluster. This is expected, as clusters with masses lower than $10^4 \, M_\odot$ are completely tidally disrupted by the end of their simulations, with most mergers occurring in the tidal tails during their lifetime.

We find that the WD–WD merger efficiencies for the $10^3$ and $10^4 \, M_\odot$ cluster simulations are nearly 3 times higher than the $10^2 \, M_\odot$ clusters. However, there is only a slight increase in WD–WD merger efficiencies between the $10^3$ and $10^4 \, M_\odot$ cluster cases. Furthermore, when we average over all cluster masses and compare the WD–WD merger rate to the merger rate from isolated BSE, there is a slight decrease in WD–WD mergers. This may suggest that WD–WD mergers are slightly suppressed by cluster dynamics. Massive stars and binaries sink to the cluster center due to mass segregation, and binary–binary or binary–single star encounters can excite the eccentricities and break up WD–WDs or their progenitor binaries. For example, some WD–WD progenitors attain high eccentricities through dynamical interactions, and merge before they evolve to WD–WDs.

As shown in Figure 3, we also find that dynamical interactions in star clusters affect the distribution of WD–WD

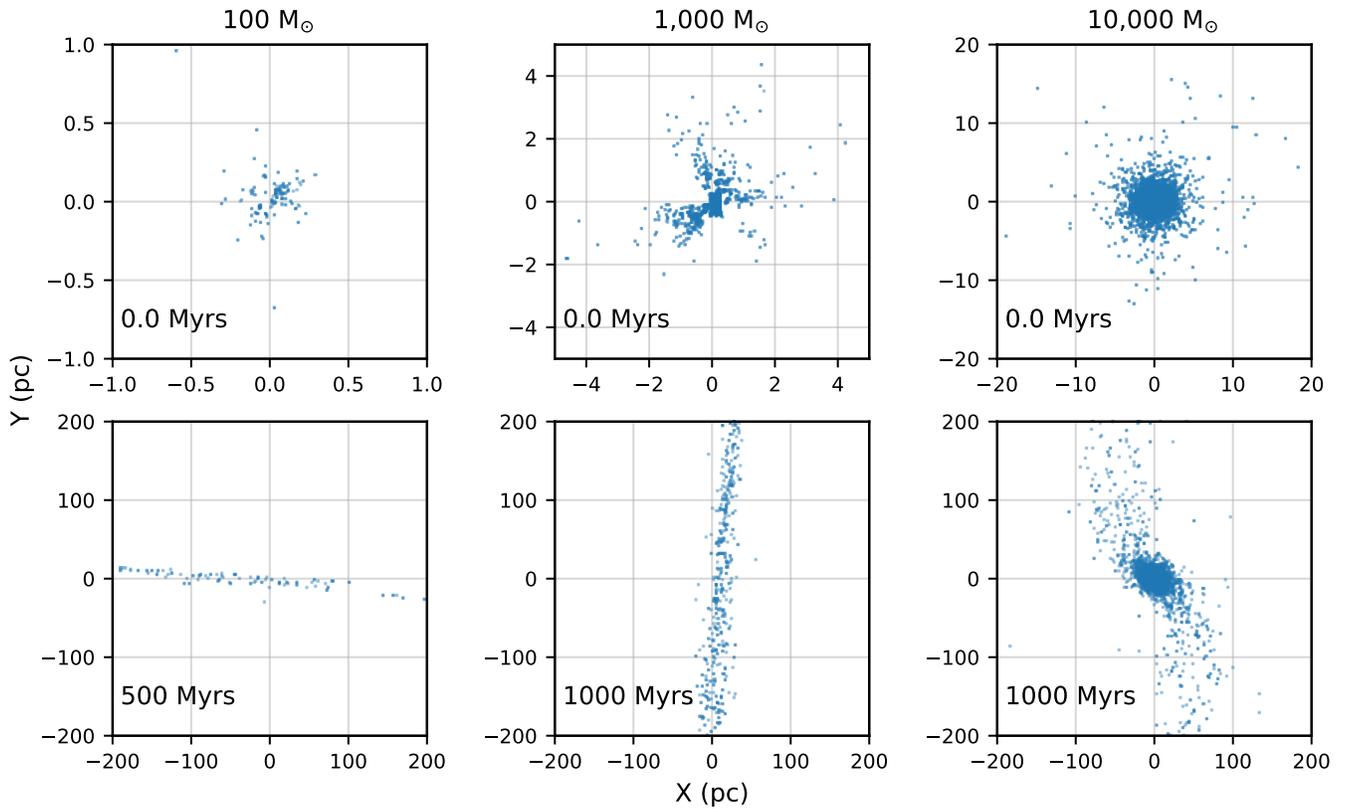

**Figure 2.** Cluster status at the beginning (top) and end (bottom) of the simulation. Going left to right, plotted are examples of our $10^2$, $10^3$, and $10^4\,M_\odot$ clusters. Each point represents a stellar object in the system.

merger time in the lower-mass clusters of $10^2$ and $10^3\,M_\odot$. Cluster dynamics prolong the merger-time distribution when compared to isolated BSE. On the other hand, there is no clear difference in the merger time for $10^4\,M_\odot$ clusters. This is likely because of the initial conditions. As shown in Table 1, the relaxation time of the $10^4\,M_\odot$ cluster is more than 30 Myr. In addition, fractal initial conditions immediately collapse locally, and stars experience dynamical interactions earlier compared to spherical initial conditions. Fractalization in the lower-mass clusters can allow for pockets of locally clustered stars, but with a spherical distribution the dynamic activity is more limited to the center of the cluster. The merger eccentricities are further evidence that WD–WD binaries are affected by the dynamical interactions inside our $10^2$ and $10^3\,M_\odot$ star clusters. Without dynamical interactions (BSE runs), the eccentricities of WD–WD mergers are all circular ($\lesssim 10^{-3}$). This is the same for the $10^4\,M_\odot$ clusters. In contrast, the $10^2$ and $10^3\,M_\odot$ clusters have a slightly higher eccentricity distribution than the $10^4\,M_\odot$ clusters.

Although the merger-time and mass distributions of WD–WD mergers do not largely differ from isolated binary evolution in the $10^4\,M_\odot$ cluster model, we observe dynamical interactions in these higher-mass clusters as well. We find a ∼10% decrease of the WD–WD mergers in these clusters (see Table 2). We further find that the progenitor binaries are dynamically perturbed by interactions inside the star clusters and evolved to some other compact objects. Unlike less massive clusters, they mostly merge inside star clusters because of the longer survival time of the host clusters.

Lastly, it should also be noted that the higher total mass fraction for WD–WD mergers within the $10^2\,M_\odot$ clusters is due to the nature of our simulations by stopping the simulations for these lower-mass clusters at 500 Myr. Later-time, lower-mass WD–WD mergers begin to dominate the distribution (for example, see Figure 4 in Section 3.4).

*3.2.2. WD–NS*

Except for our $10^3\,M_\odot$ clusters and $10^3$ and $10^4\,M_\odot$ BSE models, the numbers of mergers are less than 10, which is too small to discuss the significance of their distributions. However, the merger efficiencies of cluster runs are lower than those of isolated BSE runs for all cluster masses (see Table 2). Though our sample size is small, similar to WD–WD mergers, this is likely due to dynamical interactions that break up binaries or excite their eccentricity and cause them to merge earlier in their stellar evolution. In terms of dynamically formed binaries, our $10^2$ and $10^4\,M_\odot$ clusters produce no mergers that were not already primordial binaries, whereas in our $10^3\,M_\odot$ clusters, two out of 24 total mergers are from nonprimordial binaries.

Merger pairs also seem to be affected by the dynamical interaction inside star clusters, though they finally merge outside the cluster as seen by Figure 3. When looking at the $10^3\,M_\odot$ cluster model, binaries with smaller total mass ($M_1 + M_2$) tend to decrease due to the dynamical interactions (the number of WD–NS samples is not enough to comment on the other two cluster models). There is also a delay in merger time for WD–NS mergers when compared to isolated BSE. Our simulations may suggest that cluster dynamics can help prolong WD–WD and WD–NS merger timescales, providing an avenue in the future to explore ways to generate various transient events and young neutron stars in older open clusters.

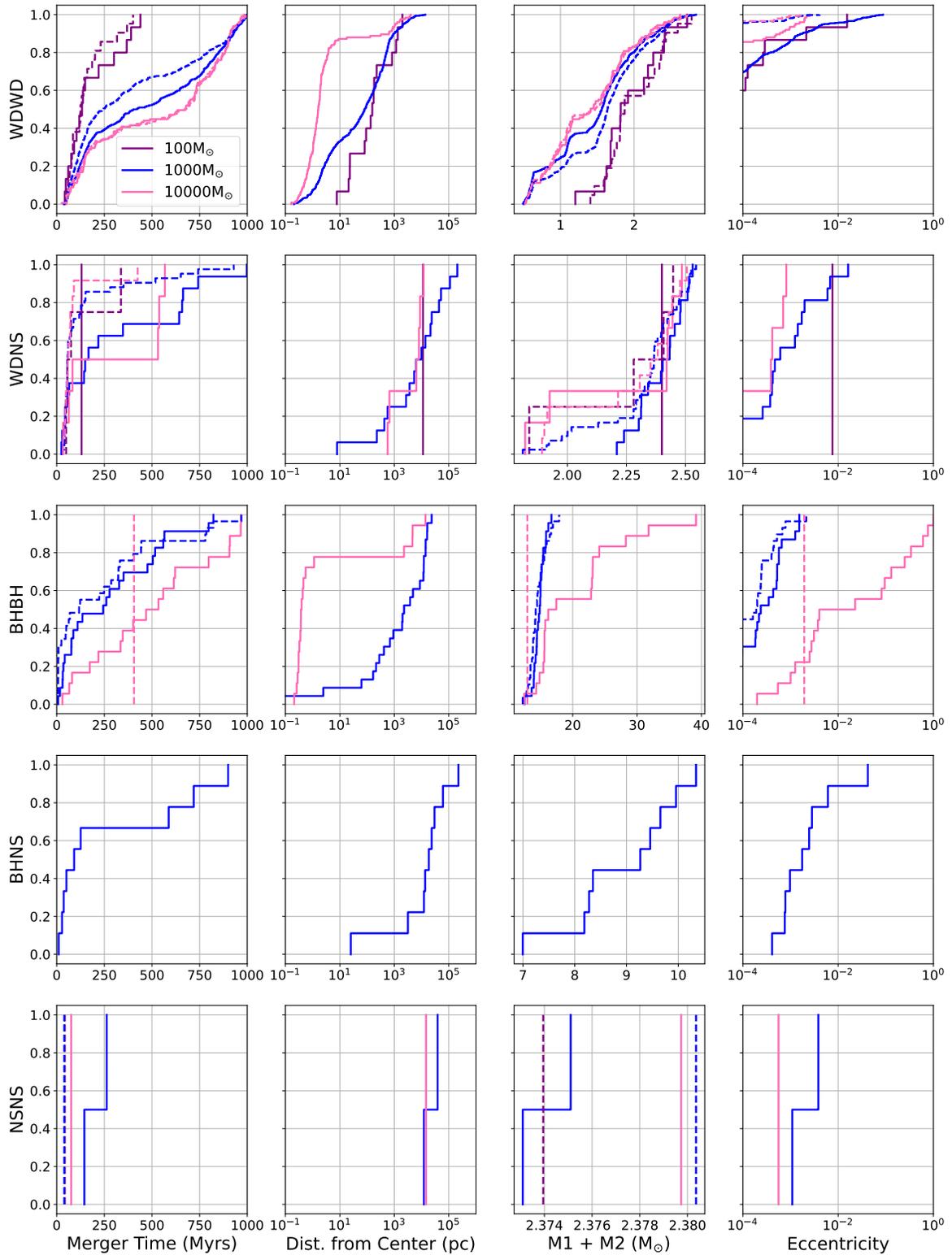

**Figure 3.** Merger time, distance from cluster center, sum of the component masses, and eccentricity distribution for each merger type right before merging. Solid lines indicate mergers from our simulations with full cluster dynamics, and dashed lines indicate mergers from our isolated BSE runs. There is no distance information for the BSE mergers. Because of the numerical accuracy, we set the minimum eccentricity for the plot to $\sim 10^{-4}$. Note that all the WD–NS mergers in the BSE runs have an eccentricity of $<10^{-4}$. Vertical lines indicate that only one event was present for the respective merger type.

### 3.2.3. BH–BH Mergers

BH–BH merger efficiencies have been examined in previous studies for open clusters (e.g., U. N. Di Carlo et al. 2019; J. Kumamoto et al. 2019; S. Rastello et al. 2019). Similar to the previous studies, the dynamical BH–BH merger efficiency increases when the cluster mass becomes massive enough. The border cluster mass seems to be between $10^3$ and $10^4\,M_\odot$. For $10^4\,M_\odot$ clusters, there is an approximately 18 times increase in

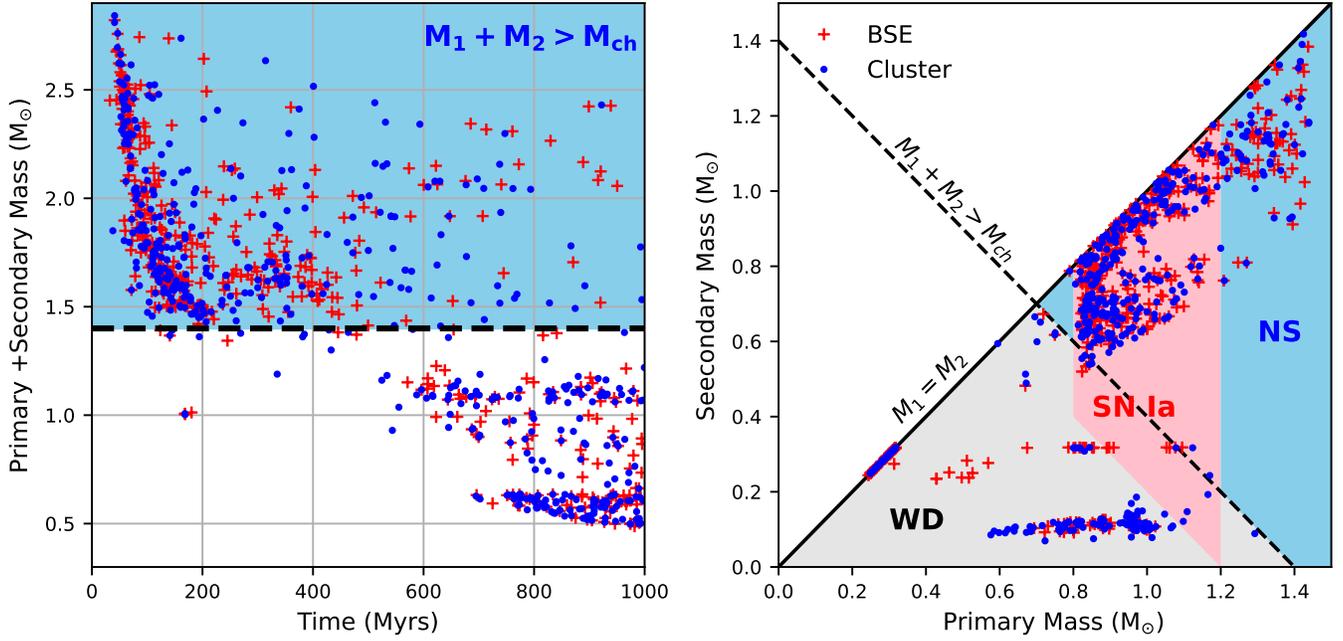

**Figure 4.** Left: sum of WD–WD merger component masses over merger time. Right: WD–WD merger mass-weighted outcome. Plotted are WD–WD mergers for $10^3\,M_\odot$ as an example, as other cluster masses follow similar distributions. Blue points are mergers from the cluster simulations, and red points are from our isolated BSE runs. The dashed line on both plots represents the Chandrasekhar limit. The shaded regions on the right plot indicate a simplified version of which WD–WD mergers we expect to result in WD, NS, and Type Ia supernovae, adopted from K. Kremer et al. (2023) and K. J. Shen (2015).

**Table 2**
Merger Efficiencies of Each Compact Object Binary per Solar Mass from Our Simulations

| Type | Cluster Efficiencies ($M_\odot^{-1}$) | | | BSE Efficiencies ($M_\odot^{-1}$) | | | Cluster Rates (Gpc$^{-3}$ yr$^{-1}$) | | |
|---|---|---|---|---|---|---|---|---|---|
| | $10^2\,M_\odot$ | $10^3\,M_\odot$ | $10^4\,M_\odot$ | $10^2\,M_\odot$ | $10^3\,M_\odot$ | $10^4\,M_\odot$ | $10^2\,M_\odot$ | $10^3\,M_\odot$ | $10^4\,M_\odot$ |
| WD–WD | $2.5 \times 10^{-4}$ | $7.0 \times 10^{-4}$ | $7.4 \times 10^{-4}$ | $3.5 \times 10^{-4}$ | $6.6 \times 10^{-4}$ | $8.6 \times 10^{-4}$ | 46–460 | 130–1300 | 140–1400 |
| WD–NS | $(1.7 \times 10^{-5})$ | $2.2 \times 10^{-5}$ | $(3.5 \times 10^{-5})$ | $(6.7 \times 10^{-5})$ | $5.8 \times 10^{-5}$ | $7.1 \times 10^{-5}$ | $(3.1–31)$ | 4.1–41 | $(6.5–65)$ |
| BH–BH | 0.0 | $3.2 \times 10^{-5}$ | $1.1 \times 10^{-4}$ | 0.0 | $4.0 \times 10^{-5}$ | $(5.9 \times 10^{-6})$ | 0.0 | 5.8–58 | 20–200 |
| BH–NS | 0.0 | $(1.2 \times 10^{-5})$ | 0.0 | 0.0 | 0.0 | 0.0 | 0.0 | $(2.3–23)$ | 0.0 |
| NS–NS | 0.0 | $(2.7 \times 10^{-6})$ | $(5.9 \times 10^{-6})$ | $(1.7 \times 10^{-5})$ | $(1.4 \times 10^{-6})$ | 0.0 | 0.0 | $(0.51–5.1)$ | $(1.1–11)$ |

**Note.** Merger rates are calculated using the cluster efficiencies and Equation (2); lower limits are calculated using a cluster formation efficiency $f_{\rm cluster} = 0.1$, and upper limits are calculated using $f_{\rm cluster} = 1$. Brackets indicate that the number of mergers in the sample is less than 10.

BH–BH mergers when compared to the isolated BSE case. As expected from previous studies of BH–BH dynamics in star clusters (e.g., S. F. Portegies Zwart & S. L. W. McMillan 2000), our BH–BH mergers in the $10^4\,M_\odot$ clusters are enhanced by dynamical formations in the dense cluster cores. Whereas the BH–BH mergers in $10^3\,M_\odot$ clusters are all from primordial binaries, half of the BH–BH mergers in $10^4\,M_\odot$ clusters are nonprimordial and formed through dynamical interactions (exchanges) in the host cluster. J. Kumamoto et al. (2019) performed similar N-body simulations for $2.5 \times 10^3$ and $10^4\,M_\odot$ clusters without primordial binaries, and obtained a merger efficiency of $4.0 \times 10^{-5}$ and $1.7 \times 10^{-5}$, respectively, in contrast to our results. Our BH–BH merger efficiencies for the $10^4\,M_\odot$ clusters are nearly 10 times larger than these results, though this can be understood by our initial binary fraction of 100%.

The dynamical formation of binary black holes (BBHs) in $10^4\,M_\odot$ clusters is clear if we see the fraction of BHs in the cluster core. In Figure 7 in Appendix B, we present the number of BHs in the core per initial cluster mass (similar to an efficiency). The fraction of BHs in $10^4\,M_\odot$ clusters is higher than that in $10^3\,M_\odot$ clusters. BBHs interact with other stars and BHs, are tightened by dynamical interactions, and merge. As shown in Figure 3, the majority (~80%) of BH–BH mergers occurred inside the cluster for the $10^4\,M_\odot$ model.

About 30% of BH–BHs merge with eccentricities >0.1 for $10^4\,M_\odot$ clusters. Out of the six eccentric BH–BH mergers, only one was formed from a primordial binary. Thus, they are mainly dynamically formed binaries. Eccentric BH–BH mergers have theoretically been suggested (J. Samsing & E. Ramirez-Ruiz 2017), and the occurrence rate has been estimated with numerical simulations of more massive ($\gtrsim 10^5\,M_\odot$) clusters such as globular clusters (C. L. Rodriguez & A. Loeb 2018; J. Samsing et al. 2018), active galactic nuclei (H. Tagawa et al. 2021), first star clusters (L. Wang et al. 2022), and young massive clusters (M. Dall'Amico et al. 2024). Our results show that even $\sim 10^4\,M_\odot$ clusters with solar metallicity contribute to eccentric BH–BH mergers.

Figure 3 also indicates the absence of eccentric BH–BH mergers for $10^3\,M_\odot$ clusters. All of their eccentricities are smaller than 0.01. BH–BH mergers for $10^3\,M_\odot$ clusters originate from primordial binaries and happen outside the clusters. In other words, they merge through isolated binary evolution, not through dynamical interactions. The absence of eccentric BH–BH mergers is also seen in previous simulations for $10^3$–$10^4 M_\odot$ star clusters (U. N. Di Carlo et al. 2019; J. Kumamoto et al. 2019). Meanwhile, S. Rastello et al. (2019) performed simulations of $3 \times 10^2 \sim 10^3 M_\odot$ clusters and reported some highly eccentric cases, though they initially included BH–BH binaries in the cluster center. Our results instead clearly show that there is a boundary between the presence and absence of eccentric BH–BH mergers, or between dynamically active and inactive clusters in the cluster mass range from $10^3$ to $10^4 M_\odot$.

### 3.2.4. BH–NS mergers

In the $10^3 M_\odot$ clusters, we obtained several BH–NS mergers which were not seen in the $10^2$ and $10^4 M_\odot$ clusters. Four out of nine of the total BH–NS mergers are dynamical, showing that $10^3 M_\odot$ clusters are able to dynamically form BH–NS binaries early in their evolution, to eventually merge. The BH–NS binary mergers occur in the outside (tidal tail) of the $10^3 M_\odot$ clusters, and there are two main time periods when these events occur. We see these mergers occurring at early times of $\lesssim 150$ Myr, with a gap in merger activity until $\gtrsim 600$ Myr. However, the sample size is too small to comment on the significance (<10), and this gap may be due to a low number of mergers.

Although the number of samples is small, we can calculate the efficiency to be $1.2 \times 10^{-5} M_\odot^{-1}$. This value is higher than a previous study for similar cluster mass and metallicity, $1 \times 10^{-7} M_\odot^{-1}$ for solar metallicity and $2 \times 10^{-6} M_\odot^{-1}$ for subsolar metallicity (S. Rastello et al. 2019). This discrepancy may be due to differences in the initial conditions and stellar evolution model. Similar to our result, they also found an increase in BH–NS merger efficiency due to dynamical interactions inside star clusters for the solar-metallicity case.

The lack of BH–NS mergers in our high-mass clusters specifically agrees with previous studies such as G. Fragione & S. Banerjee (2020). BHs are more massive than NSs. Therefore, they segregate in the cluster center, dynamically heat up the core, and prevent the formation of BH–NS binaries. As shown in Figure 7 in Appendix B, the number of BH–NS binaries inside clusters for $10^4 M_\odot$ clusters is much smaller than that for $10^3 M_\odot$ clusters. Similarly, the cluster cores of $10^4 M_\odot$ clusters are dominated by BHs.

### 3.3. Compact Binary Merger Rates

In this section, we focus on WD–WD, WD–NS, NS–NS, BH–BH, and BH–NS merger rates. We adopt a cosmic star formation rate density as a function of redshift from P. Madau & T. Fragos (2017):

$$\psi(z) = 0.01 \frac{(1+z)^{2.6}}{1 + [(1+z)/3.2]^{6.2}} \, M_\odot \, \text{yr}^{-1} \, \text{Mpc}^{-3}, \quad (1)$$

which assumes a P. Kroupa (2001) IMF. We assume a certain fraction of all stars are formed in a cluster, $f_{\text{cluster}}$, giving us a rate density equation of

$$\mathfrak{R}(z) = \eta \times \psi(z) \times f_{\text{cluster}}, \quad (2)$$

where $\eta$ is the merger efficiency, and $\mathfrak{R}(z)$ is the cosmic rate density as a function of redshift. For a Milky Way–like galaxy, $f_{\text{cluster}}$ is $\approx 0.1$ (A. Misiriotis et al. 2006; A. E. Piskunov et al. 2007; N. Bastian 2008). However, in order to compare to past studies, we also compute an upper limit of the rate assuming an $f_{\text{cluster}}$ of 1. We report our rates as a range, using $f_{\text{cluster}} = 0.1$ and $f_{\text{cluster}} = 1$ as the lower and upper bounds, respectively. Table 2 reports on the rate density for the local Universe.

In addition to providing separate merger rates for each cluster mass, we calculate a singular merger rate for the local Universe. We take a mass-dependent cluster formation efficiency $f_{\text{cluster,M}} \propto M^{-1}$ from a cluster mass function of $dN/dM \propto M^{-2}$ (S. F. Portegies Zwart et al. 2010), and assume all stars are formed in clusters of masses $10^2$, $10^3$, and $10^4 M_\odot$ such that $f_{\text{cluster},100} + f_{\text{cluster},1000} + f_{\text{cluster},10000} = 1$. Here, $f_{\text{cluster},100}$, $f_{\text{cluster},1000}$, and $f_{\text{cluster},10000}$ are the cluster formation efficiencies for $10^2$, $10^3$, and $10^4 M_\odot$ clusters, respectively. These rate calculations will be referred to as $\mathfrak{R}_{\text{Local}}$ in the following text.

We find a $\mathfrak{R}_{\text{BHNS}}$ of 2.3–23 Gpc$^{-3}$ yr$^{-1}$ for clusters of mass $10^3 M_\odot$; this is in rough agreement with past studies on open clusters and young massive clusters (S. Rastello et al. 2020; M. Arca Sedda et al. 2024), and a couple magnitudes larger than recent estimates of BH–NS mergers in globular clusters (e.g., C. S. Ye et al. 2019). This range is also in rough agreement with the limits as calculated by LIGO-Virgo observations (7.8–140 Gpc$^{-3}$ yr$^{-1}$; R. Abbott et al. 2023b), and further confirms that open clusters contribute a significant fraction of the overall BH–NS merger rate. If we instead assume a mass-dependent cluster efficiency for the local Universe, the rate drops to $\mathfrak{R}_{\text{Local}} = 2.1$ Gpc$^{-3}$ yr$^{-1}$, and is within the LIGO-Virgo observational limits.

We find a $\mathfrak{R}_{\text{BHBH}}$ of 5.8–58 Gpc$^{-3}$ yr$^{-1}$ for clusters of mass $10^3 M_\odot$, and 20–200 Gpc$^{-3}$ yr$^{-1}$ for clusters of mass $10^4 M_\odot$. These values are similar, within an order of magnitude, to other simulations of open clusters and young massive clusters (U. N. Di Carlo et al. 2020; J. Kumamoto et al. 2020; F. Santoliquido et al. 2020; M. Arca Sedda et al. 2024). We further find our rate calculated from $10^3 M_\odot$ clusters is similar to the local merger rate found in globular cluster studies (e.g., A. Askar et al. 2016; C. L. Rodriguez et al. 2016; C. L. Rodriguez & A. Loeb 2018), showing that open clusters contribute to overall BH–BH merger rates. However, we find that for high-mass clusters, our $\mathfrak{R}_{\text{BHBH}}$ range is larger than the upper limit for BH–BH mergers as provided by LIGO-Virgo observations, which is 17.9–44 Gpc$^{-3}$ yr$^{-1}$ (R. Abbott et al. 2023b). If we instead assume a mass-dependent cluster efficiency for the local Universe, the total BH–BH merger rate across all cluster masses is $\mathfrak{R}_{\text{Local}} = 7.0$ Gpc$^{-3}$ yr$^{-1}$. This rate is well within observational limits set by LIGO-Virgo.

We further calculate the merger rates for our eccentric BH–BH mergers in the previous subsection. The $10^4 M_\odot$ clusters produce eccentric BH–BH merger rates of 6.5–65 Gpc$^{-3}$ yr$^{-1}$.

As for NS–NS mergers, we find a $\mathfrak{R}_{\text{NSNS}}$ of 0.51–5.1 Gpc$^{-3}$ yr$^{-1}$ for clusters of mass $10^3 M_\odot$, and 1.1–11 Gpc$^{-3}$ yr$^{-1}$ for clusters of mass $10^4 M_\odot$ (though keep in mind our sample sizes are small). Our mass-dependent rate across all clusters is $\mathfrak{R}_{\text{Local}} = 0.56$ Gpc$^{-3}$ yr$^{-1}$. LIGO-Virgo calculated upper and lower limits of 10–1700 Gpc$^{-3}$ yr$^{-1}$ for the rates of NS–NS mergers (R. Abbott et al. 2023b). That being said, past studies of young and open clusters have reported rates as small as 0.01–1 Gpc$^{-3}$ yr$^{-1}$ (G. Fragione & S. Banerjee 2020) and as large as 151 Gpc$^{-3}$ yr$^{-1}$ (F. Santoliquido et al. 2020). Meanwhile, simulations of globular clusters have suggested a local NS–NS merger rate of 0.02 Gpc$^{-3}$ yr$^{-1}$ (C. S. Ye et al. 2019). Although our range of rates mostly falls below the

Table 3
Efficiencies of NS Formation and Type Ia Supernovae from Our Simulations, Given by Figure 4

| Remnant Type | Cluster Efficiencies ($M_\odot^{-1}$) | | | BSE Efficiencies ($M_\odot^{-1}$) | | |
|---|---|---|---|---|---|---|
| | $10^2\,M_\odot$ | $10^3\,M_\odot$ | $10^4\,M_\odot$ | $10^2\,M_\odot$ | $10^3\,M_\odot$ | $10^4\,M_\odot$ |
| NS | $2.3 \times 10^{-4}$ | $4.2 \times 10^{-4}$ | $3.8 \times 10^{-4}$ | $3.5 \times 10^{-4}$ | $4.7 \times 10^{-4}$ | $4.5 \times 10^{-4}$ |
| Type Ia SN | $1.8 \times 10^{-5}$ | $3.5 \times 10^{-5}$ | $2.8 \times 10^{-5}$ | $3.0 \times 10^{-5}$ | $4.2 \times 10^{-5}$ | $3.9 \times 10^{-5}$ |

**Note.** Note that the integration time is 500 Myr for the $10^2\,M_\odot$ clusters, but 1000 Myr for the other clusters.

LIGO-Virgo range, open clusters could still contribute a nonnegligible fraction to the overall NS–NS merger rates.

As for WD–WD mergers, we find a rate of 46–460, 130–1300, and 140–1400 Gpc$^{-3}$ yr$^{-1}$ for our low-, intermediate-, and high-mass clusters, respectively. We further calculate $\mathfrak{R}_{\rm Local} = 540$ Gpc$^{-3}$ yr$^{-1}$. These rates are larger than what was previously calculated by globular clusters in the local Universe (D. Liu & B. Wang 2020; K. Kremer et al. 2021b, 2023). As for WD–NS mergers, we find a rate $\mathfrak{R}_{\rm WDNS}$ of 3.1–31, 4.1–41, and 6.5–65 Gpc$^{-3}$ yr$^{-1}$ for our low-, intermediate-, and high-mass clusters, respectively, and $\mathfrak{R}_{\rm Local}$ is calculated to be 32 Gpc$^{-3}$ yr$^{-1}$. Our rates and efficiencies for WD–NS mergers are similar to calculations through previous population synthesis models (S. Toonen et al. 2018).

We note that our rates, and thus conclusions, are highly limited by our sample sizes; this is especially true for mergers that only have a few events, such as the NS–NS mergers. In addition, our three separate cluster cases vary in total mass when summed over all clusters; the $10^3\,M_\odot$ clusters consist of a total mass ∼4 times greater than the $10^4\,M_\odot$ clusters, and ∼12 times greater than the $10^2\,M_\odot$ clusters. In order to have a more accurate understanding of such mergers, more simulations need to be run in the future. Furthermore, it is important to acknowledge that the $10^2\,M_\odot$ merger rates are biased by an incomplete IMF (on a cluster-by-cluster basis) due to their low mass.

### 3.4. WD–WD Merger Remnants

#### 3.4.1. Implications for Fast Radio Bursts

One of the leading theories for the creation of FRBs is via magnetars (Y.-P. Yang & B. Zhang 2018; CHIME/FRB Collaboration et al. 2020; B. Zhang 2020), especially since one Galactic magnetar has shown FRB-like radio emission (SGR 1935+2154; C. D. Bochenek et al. 2020; CHIME/FRB Collaboration et al. 2020). These young neutron stars are more likely to arise from core-collapse supernovae in young stellar populations. However, there has been at least one FRB associated with a globular cluster in M81 (M. Bhardwaj et al. 2021; F. Kirsten et al. 2022), and this FRB may be formed via a WD–WD merger (K. Kremer et al. 2021b). If the total mass of the merging WDs exceeds the Chandrasekhar limit, such remnant may form a NS (K. Nomoto & I. Iben 1985; J. Schwab 2021). In fact, previous studies of FRB formation in globular clusters have argued that WD–WD mergers are the most plausible formation of young neutron stars in clusters (W. Lu et al. 2021; K. Kremer et al. 2021a, 2023). Here, we investigate the WD–WD merger avenue for young neutron star formation in open clusters.

The left panel of Figure 4 shows the component mass distribution over merger time for all WD–WD events. For all cluster masses, most super-Chandrasekhar mergers occur in the beginning of the clusters' lifetimes, and after around 500 Myr (the simulation time for the $10^2\,M_\odot$ clusters), most of the mergers are below the Chandrasekhar mass limit. We calculate the formation efficiency of super-Chandrasekhar mergers before 500 Myr to be $2.3 \times 10^{-4}$, $3.5 \times 10^{-4}$, and $3.1 \times 10^{-4}\,M_\odot^{-1}$, for the $10^2$, $10^3$, and $10^4\,M_\odot$ clusters, respectively. We calculate the formation efficiency of super-Chandrasekhar mergers after 500 Myr to be $7.1 \times 10^{-5}$ and $6.5 \times 10^{-5}\,M_\odot^{-1}$, for clusters of mass $10^3$ and $10^4\,M_\odot$, respectively. We estimate the total super-Chandrasekhar WD–WD merger rate range in the local Universe to be 78–780 and 70–700 Gpc$^{-3}$ yr$^{-1}$ for the $10^3$ and $10^4\,M_\odot$ clusters, respectively.

With this said, not all super-Chandrasekhar WD–WD mergers are thought to form NSs. The right panel of Figure 4 shows the detailed primary and secondary mass distribution of WD–WD mergers obtained from our simulations. Similar to Figure 5 of K. Kremer et al. (2023), we also show the expected merger outcome from K. J. Shen (2015).

From this plot, the NS formation efficiencies during 1000 Myr are $4.2 \times 10^{-4}$ and $3.8 \times 10^{-4}\,M_\odot^{-1}$, for clusters of mass $10^3$ and $10^4\,M_\odot$, respectively. These values are summarized in Table 3. Compared to isolated BSE runs, the formation efficiencies of super-Chandrasekhar NSs are decreased across all cluster runs. We also expect the NS formation rate from WD–WD mergers to be 15–150 and 21–210 Gpc$^{-3}$ yr$^{-1}$ for $10^3$ and $10^4\,M_\odot$ clusters, respectively. For BSE, these rates are 11–110 Gpc$^{-3}$ yr$^{-1}$ for both the $10^3$ and $10^4\,M_\odot$ clusters.

Although the fraction of NSs formed from WD–WD mergers decreases in our star clusters when compared to BSE, the event rate of super-Chandrasekhar WD–WD mergers is larger than that calculated for globular clusters, ∼10 Gpc$^{-3}$ yr$^{-1}$ (K. Kremer et al. 2021a). When considering the WD–WD merger scenario to create young neutron stars, our results suggest that it might be more likely to detect an FRB from an open cluster or its remnant. Even if the cluster is possibly dissolved due to tidal stripping, the original cluster dynamics will have prolonged consequences on the stars born in the cluster; open clusters might, therefore, increase the likelihood of detecting FRBs in older stellar populations in the Milky Way and other galaxies.

#### 3.4.2. Implications for Type Ia Supernovae

Simulations suggest that a fraction of WD–WD mergers will instead result in Type Ia supernovae (e.g., R. Pakmor et al. 2013; K. J. Shen & L. Bildsten 2014), with component mass distributions in accordance with our Figure 4. As mentioned in Section 1, the merging of two carbon–oxygen WDs is believed to be a source of Type Ia supernovae, as well as direct, head-on collisions (e.g., S. Rosswog et al. 2009; B. Katz & S. Dong 2012; D. Kushnir et al. 2013). Star clusters are a possible place to generate such Type Ia supernovae, and past studies of open clusters have suggested this (M. M. Shara & J. R. Hurley 2002).

Based on our Figure 4, we can estimate which mergers we expect to result in Type Ia supernovae. From our simulations, we obtain the formation efficiency of Type Ia supernovae as $1.8 \times 10^{-5}$, $3.5 \times 10^{-5}$, $2.8 \times 10^{-5}$ for the $10^2$, $10^3$, and $10^4\,M_\odot$ clusters. Similar to super-Chandrasekhar NS formation efficiency, the Type Ia supernova formation efficiency also is decreased in star clusters when compared to BSE (see Table 3). As a result, we calculate a local Type Ia supernova rate of $34$–$340\,\mathrm{Gpc}^{-3}\,\mathrm{yr}^{-1}$ for clusters of mass $10^2\,M_\odot$, $64$–$640\,\mathrm{Gpc}^{-3}\,\mathrm{yr}^{-1}$ for clusters of mass $10^3\,M_\odot$, and $51$–$510\,\mathrm{Gpc}^{-3}\,\mathrm{yr}^{-1}$ for clusters of mass $10^4\,M_\odot$. To compare, the observed rate of Type Ia supernovae in the local Universe is $(2.5 \pm 0.5) \times 10^4\,\mathrm{Gpc}^{-3}\,\mathrm{yr}^{-1}$ (W. Li et al. 2011; E. Cappellaro et al. 2015). Our simulations of open clusters only account for 0.14%–2.6% of the total observed Type Ia supernova rate in the local Universe. These values are similar to the percentages found in simulations of globular clusters from K. Kremer et al. (2021b).

## 4. Conclusions

In this paper, we performed a series of $N$-body simulations of $10^2$, $10^3$, and $10^4\,M_\odot$ open clusters with a single and binary stellar evolution model. For comparison, we also performed isolated binary evolution simulations for exactly the same binaries evolved in our star cluster simulations.

Relatively massive open clusters with $10^4\,M_\odot$, modeling young massive clusters, survive for our simulation time (1000 Myr). Their evolution is similar to globular clusters, and BH–BH binaries are dynamically formed in their cores. As a result, BH–BH merger efficiency significantly increases in the clusters. Some of them are eccentric. Instead, WD–WD and WD–NS merger rates decrease in the cluster models when compared to isolated binary stellar evolution.

In star clusters with a mass typical for open clusters ($10^2$ and $10^3\,M_\odot$), BH–BH mergers are suppressed. Compared to isolated binary stellar evolution, WD–WD merger efficiencies increase in the $10^3\,M_\odot$ clusters but decrease in the $10^2\,M_\odot$ clusters. WD–NS mergers decrease in both the $10^2$ and $10^3\,M_\odot$ clusters. Furthermore, we found BH–NS mergers only in the $10^3\,M_\odot$ clusters, although the number of samples is not enough for further discussion.

Our overall volumetric rate for compact binary mergers in the local Universe is greater in our open cluster simulations compared to those of globular clusters. When calculating compact binary mergers that can be detected via gravitational waves by LIGO-Virgo-KAGRA, or even LISA, it is important to consider the dynamics of open clusters.

We found a fraction of super-Chandrasekhar WD–WD mergers, which may account for young magnetars and thus FRBs. The merger rate density of WD–WD mergers that collapse into a NS was found to be $11$–$210\,\mathrm{Gpc}^{-3}\,\mathrm{yr}^{-1}$, which is larger than that estimated for globular clusters. Our open clusters further account for at most 2.6% of the Type Ia supernova rate in the local Universe.

We note that our conclusions are limited by the sample sizes of our clusters. Notwithstanding, this study further shows that open clusters account for a nonnegligible amount to the compact binary populations in galaxies. With there being several more of them when compared to globular clusters, open clusters prove to be an integral part of the stellar evolution and compact merger histories in galaxies.

## Acknowledgments

S.C. was supported by the Fulbright U.S. Student Program, which is sponsored by the U.S. Department of State and Japan-U.S. Educational Commission. Its contents are solely the responsibility of the author and do not necessarily represent the official views of the Fulbright Program, the Government of the United States, or the Japan-U.S. Educational Commission. Numerical computations were in part carried out on Cray XC50 at the Center for Computational Astrophysics, National Astronomical Observatory of Japan. L.W. thanks the support from National Natural Science Foundation of China through grant Nos. 21BAA00619 and 12233013, the Hundred-Talent Program of Sun Yat-sen University, and the Fundamental Research Funds for the Central Universities, Sun Yat-sen University (grant No. 22hytd09). This work was supported by JSPS KAKENHI grant Nos. 22H01259 and 19F19317.

*Software*: PETAR (L. Wang et al. 2020a), MCLUSTER (A. H. W. Küpper et al. 2011), (J. D. Hunter 2007), galpy (J. Bovy 2015), numpy (C. R. Harris et al. 2020), scipy (P. Virtanen et al. 2020), matplotlib (J. D. Hunter 2007).

## Appendix A
## Cluster Evolution

In this section, we discuss how initial mass and shape can affect cluster evolution and disrupt timescales. We also highlight the tidal radius evolution of our cluster populations, as calculated by PETAR.

Figure 5 shows how the tidal radius evolves over time for each of our clusters. Clusters of $10^4\,M_\odot$ do not evolve too differently from one another, and remain intact over the entire 1 Gyr simulation; they only lose $\sim 30\%$ of stars by 1 Gyr. However, there is much variation in the fate of the $10^3$ and $10^4\,M_\odot$ clusters. Some clusters disrupt within a few million years after the start of the simulation, whereas others last hundreds of million years longer. This is likely caused by our initial fractalizations, where the randomness of the fractalization may allow for pockets of gravitationally bound stars to form and dominate the shape of the cluster. As mentioned in Section 2, our $10^2$ and $10^3\,M_\odot$ mass clusters start with a fractal dimension of 1.6, whereas our $10^4\,M_\odot$ mass clusters start with a fractal dimension of 3.0 (i.e., spherical).

Figure 6 shows three examples of fractalized clusters generated from different random seeds for the initial condition generator MCLUSTER, and the different evolutions of their respective tidal radii. Thus, the initial shapes of the $10^2$ and $10^3\,M_\odot$ mass clusters should contribute to how quickly the cluster is dissolved. Even for clusters of the same mass, their evolution timescales strongly vary. Further information on fractal profiles is discussed in A. H. W. Küpper et al. (2011).

Furthermore, all of our $10^2\,M_\odot$ mass clusters are completely dissolved by 500 Myr, while all of our $10^3\,M_\odot$ mass clusters are dissolved by 1 Gyr. Although some clusters appear to have a greater than zero tidal radius by the end of their simulation, in reality the tidal radius is dictated by one or two massive stars because of the algorithm we adopt. This is shown in Figure 5, where the number of stars within the tidal radius suddenly drops to or near zero.

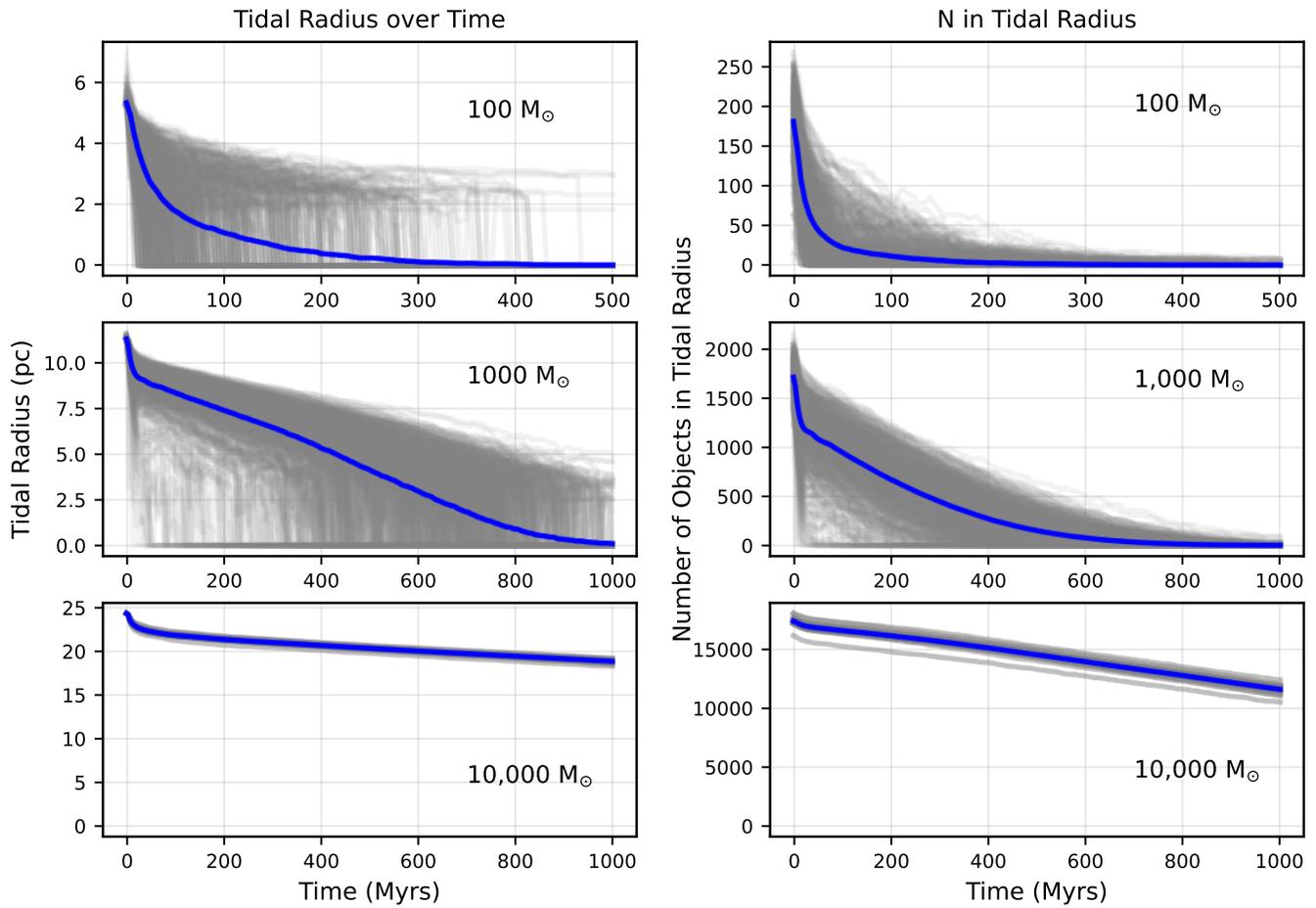

**Figure 5.** Tidal radius over simulation time. On the left, we show the size of the tidal radius of our clusters over time. Each gray line represents a different cluster, where the blue line represents the average radius at a given time. Similarly, on the left we show the number of objects within the tidal radius of our clusters.

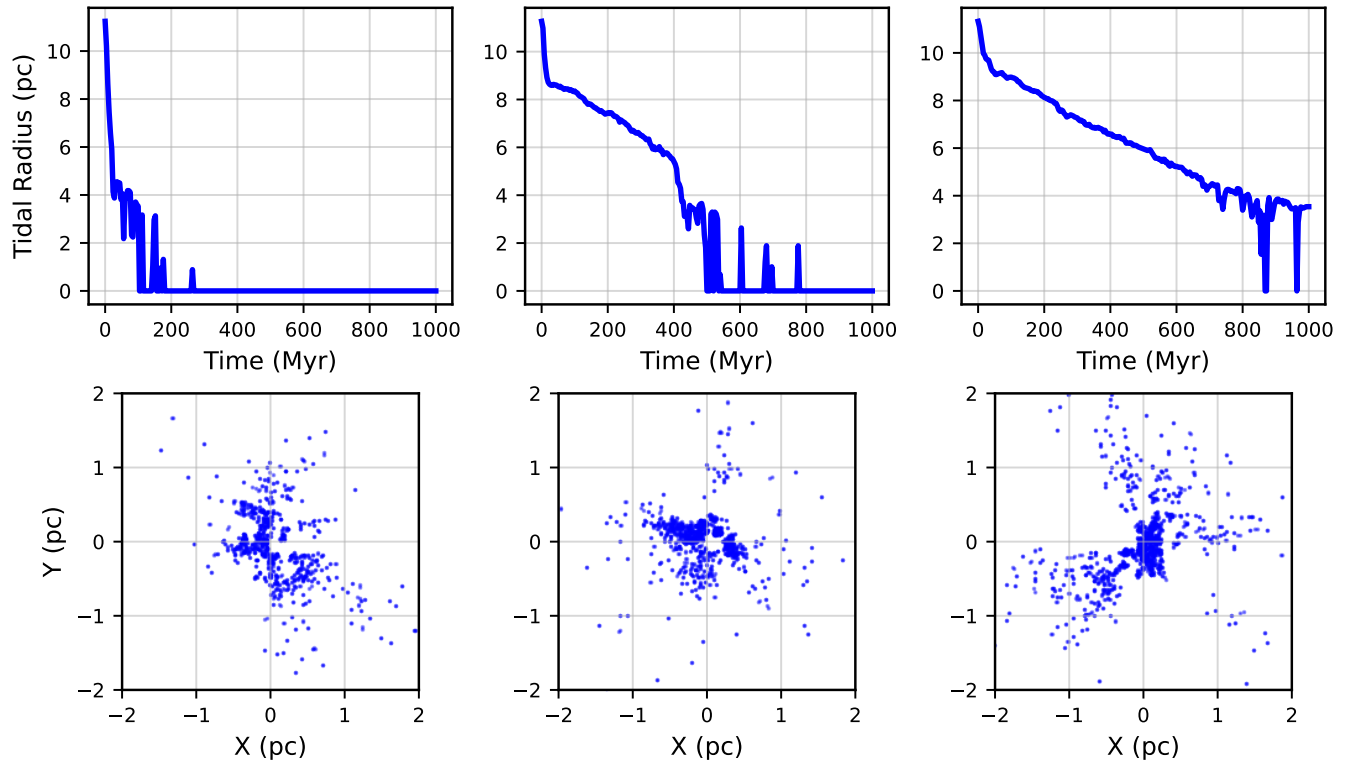

**Figure 6.** Examples of $10^3\,M_\odot$ clusters with fractal dimension 1.6. On the bottom are three examples of what a cluster with a fractal dimension of 1.6 may look like. The tidal radius over the elapsed simulation time is plotted above each respective cluster.

## Appendix B
## Binary Counts

In order to understand merger rates, here we discuss the number of compact objects over simulation time in the cores of our clusters. The core is typically the densest region of a cluster, and the dynamical evolution of compact binaries occurs in it. PETAR is able to distinguish the core radius over time following methods in S. Casertano & P. Hut (1985). Figure 7 shows the average number of the different compact objects in the core per cluster over time for each cluster mass, as well as the number of BH–NS binaries. We do not include plots of the core evolution for $10^2 M_\odot$ clusters, as they have too few stars to reliably distinguish a core by PETAR.

As shown in Figure 7, the number of BHs inside the cores of $10^4 M_\odot$ clusters is larger than that of $10^3 M_\odot$ clusters. Mass segregation causes the BHs to sink to the center of the $10^4 M_\odot$ clusters, causing them to form BH–BH binaries. The higher fraction of BHs in the cores of $10^4 M_\odot$ clusters results in the higher BH–BH merger efficiency of $10^4 M_\odot$ clusters compared to $10^3 M_\odot$ clusters. Because BHs are heavier than NSs, the number of NSs in the cores is an order of magnitude lower than that of BHs. As a consequence, there is no chance of forming BH–NS binaries in the cores. The number of BH–NS binaries is also shown in Figure 7. The number of BH–NS binaries is higher in low-mass clusters because they are tidally stretched before BH–NS binaries sink to the cluster cores.

In both $10^3$ and $10^4 M_\odot$ clusters, the number of BHs in the cores decreases with time due to the dynamical ejection. The trend for NSs seems to be similar, but the number of NSs in the cores is small. Furthermore, after ~600 Myr, the number of compact objects in the cores begins to drop in $10^3 M_\odot$ clusters. This is due to the tidal disruption of the host clusters.

In contrast to the BHs in the cores in the $10^4 M_\odot$ clusters, WD numbers increase later due to the stellar evolution. In the $10^3 M_\odot$ clusters, the number of WDs decreases with time, but it keeps a high value until ~600 Myr; on the other hand, BHs in the cores start to decrease at ~400 Myr.

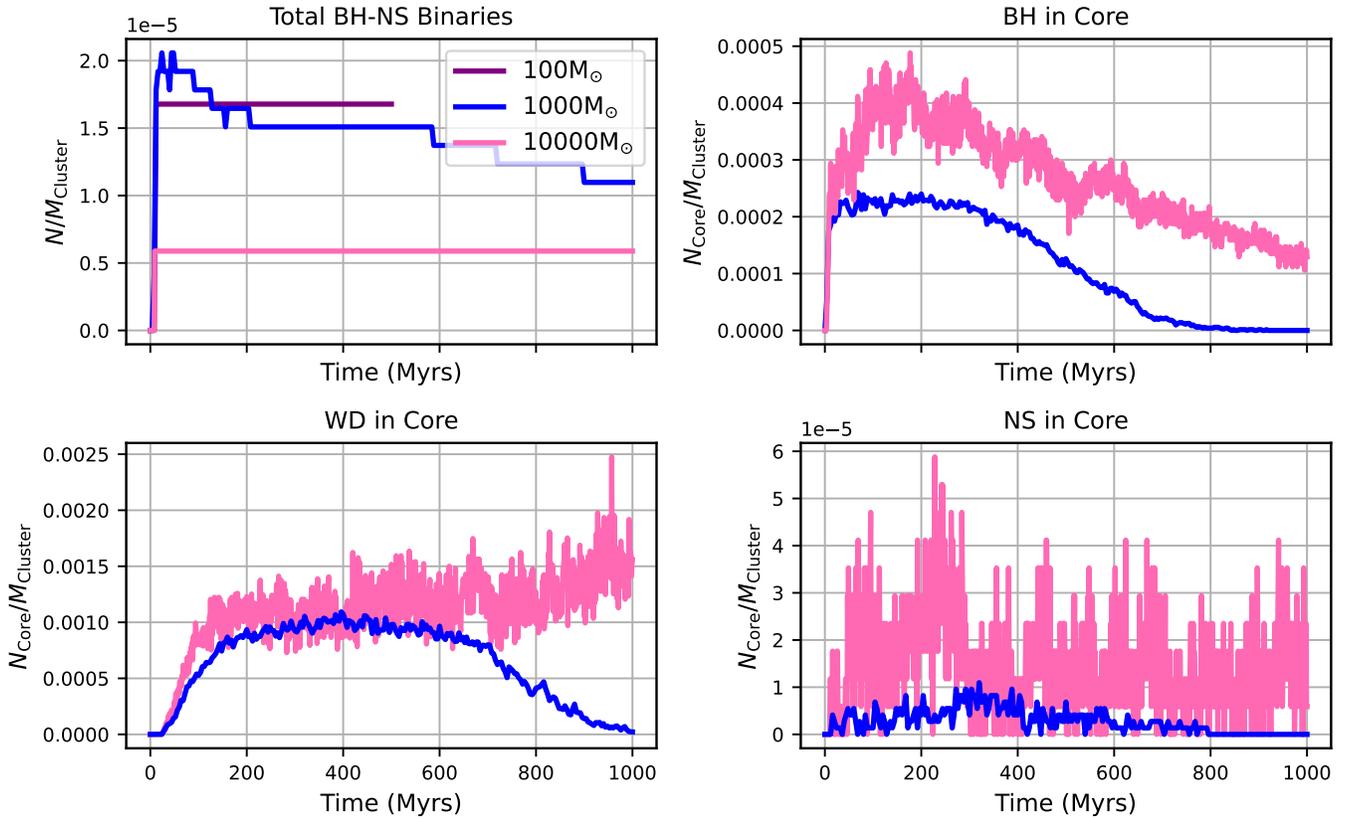

**Figure 7.** Average number of compact binaries in core over time. Counts are normalized by cluster mass, such that $N$ is summed over all simulation runs and $M_{\text{Cluster}} = N_{\text{run}} \times 10^2$, $10^3$, or $10^4 M_\odot$. We do not include the number of core BHs, NSs, or WDs in $10^2 M_\odot$ clusters. These clusters have too few stars to reliably distinguish a core by PETAR. Plotted additionally is the number of BH–NS binaries present in our simulation (not just within the tidal radius), over our simulation time for each of our clusters.

## ORCID iDs

Savannah Cary https://orcid.org/0000-0003-1860-1632
Michiko Fujii https://orcid.org/0000-0002-6465-2978
Long Wang https://orcid.org/0000-0001-8713-0366
Ataru Tanikawa https://orcid.org/0000-0002-8461-5517

## References

Abbott, R., Abbott, T. D., Acernese, F., et al. 2023a, PhRvX, 13, 041039
Abbott, R., Abbott, T. D., Acernese, F., et al. 2023b, PhRvX, 13, 011048
Amaro-Seoane, P., Andrews, J., Arca Sedda, M., et al. 2023, LRR, 26, 2
Arca Sedda, M., Kamlah, A. W. H., Spurzem, R., et al. 2024, MNRAS, 528, 5140
Askar, A., Szkudlarek, M., Gondek-Rosińska, D., Giersz, M., & Bulik, T. 2016, MNRAS, 464, L36
Banerjee, S. 2017, MNRAS, 467, 524
Banerjee, S., Belczynski, K., Fryer, C. L., et al. 2020, A&A, 639, A41
Bastian, N. 2008, MNRAS, 390, 759
Belczynski, K., Bulik, T., Fryer, C. L., et al. 2010, ApJ, 714, 1217
Belczynski, K., Heger, A., Gladysz, W., et al. 2016, A&A, 594, A97
Berger, E. 2014, ARA&A, 52, 43
Bhardwaj, M., Gaensler, B. M., Kaspi, V. M., et al. 2021, ApJL, 910, L18
Bochenek, C. D., Ravi, V., Belov, K. V., et al. 2020, Natur, 587, 59
Bovy, J. 2015, ApJS, 216, 29
Cappellaro, E., Botticella, M. T., Pignata, G., et al. 2015, A&A, 584, A62
Casertano, S., & Hut, P. 1985, ApJ, 298, 80
Cheng, S., Cummings, J. D., Ménard, B., & Toonen, S. 2020, ApJ, 891, 160
CHIME/FRB Collaboration, Andersen, B. C., Bandura, K. M., et al. 2020, Natur, 587, 54
Dall'Amico, M., Mapelli, M., Torniamenti, S., & Arca Sedda, M. 2024, A&A, 683, A186
Di Carlo, U. N., Giacobbo, N., Mapelli, M., et al. 2019, MNRAS, 487, 2947
Di Carlo, U. N., Mapelli, M., Giacobbo, N., et al. 2020, MNRAS, 498, 495
Fernández, R., & Metzger, B. D. 2013, ApJ, 763, 108
Fragione, G., & Banerjee, S. 2020, ApJL, 901, L16
Fryer, C. L., Belczynski, K., Wiktorowicz, G., et al. 2012, ApJ, 749, 91
Goodwin, S. P., & Whitworth, A. P. 2004, A&A, 413, 929
Harris, C. R., Millman, K. J., van der Walt, S. J., et al. 2020, Natur, 585, 357
Hobbs, G., Lorimer, D. R., Lyne, A. G., & Kramer, M. 2005, MNRAS, 360, 974
Hunter, J. D. 2007, CSE, 9, 90
Hurley, J. R., Pols, O. R., & Tout, C. A. 2000, MNRAS, 315, 543
Hurley, J. R., Tout, C. A., & Pols, O. R. 2002, MNRAS, 329, 897
Iben, I. J., & Tutukov, A. V. 1984, ApJS, 54, 335
Iwasawa, M., Namekata, D., Nitadori, K., et al. 2020, PASJ, 72, 13
Iwasawa, M., Oshino, S., Fujii, M. S., & Hori, Y. 2017, PASJ, 69, 81
Iwasawa, M., Tanikawa, A., Hosono, N., et al. 2016, PASJ, 68, 54
Kashiyama, K., Ioka, K., & Mészáros, P. 2013, ApJL, 776, L39
Katz, B., & Dong, S. 2012, arXiv:1211.4584
King, A., Olsson, E., & Davies, M. B. 2007, MNRAS, 374, L34
King, A. R., Pringle, J. E., & Wickramasinghe, D. T. 2001, MNRAS, 320, L45
Kirsten, F., Marcote, B., Nimmo, K., et al. 2022, Natur, 602, 585
Kremer, K., Fuller, J., Piro, A. L., & Ransom, S. M. 2023, MNRAS, 525, L22
Kremer, K., Li, D., Lu, W., Piro, A. L., & Zhang, B. 2023, ApJ, 944, 6
Kremer, K., Piro, A. L., & Li, D. 2021a, ApJL, 917, L11
Kremer, K., Rui, N. Z., Weatherford, N. C., et al. 2021b, ApJ, 917, 28
Kremer, K., Ye, C. S., Heinke, C. O., et al. 2024, ApJL, 977, L42
Kroupa, P. 1995a, MNRAS, 277, 1491
Kroupa, P. 1995b, MNRAS, 277, 1507
Kroupa, P. 2001, MNRAS, 322, 231
Kumamoto, J., Fujii, M. S., & Tanikawa, A. 2019, MNRAS, 486, 3942
Kumamoto, J., Fujii, M. S., & Tanikawa, A. 2020, MNRAS, 495, 4268
Küpper, A. H. W., Maschberger, T., Kroupa, P., & Baumgardt, H. 2011, MNRAS, 417, 2300
Kushnir, D., Katz, B., Dong, S., Livne, E., & Fernández, R. 2013, ApJL, 778, L37
Lada, C. J., & Lada, E. A. 2003, ARA&A, 41, 57
Lamberts, A., Garrison-Kimmel, S., Hopkins, P. F., et al. 2018, MNRAS, 480, 2704
Li, W., Chornock, R., Leaman, J., et al. 2011, MNRAS, 412, 1473
Liu, D., & Wang, B. 2020, MNRAS, 494, 3422
Liu, X. 2018, Ap&SS, 363, 242
Liu, X.-J., Sengar, R., Bailes, M., et al. 2025, ApJL, 981, L29
Lu, W., Beniamini, P., & Kumar, P. 2021, MNRAS, 510, 1867
Madau, P., & Fragos, T. 2017, ApJ, 840, 39
Metzger, B. D. 2012, MNRAS, 419, 827
Misiriotis, A., Xilouris, E. M., Papamastorakis, J., Boumis, P., & Goudis, C. D. 2006, A&A, 459, 113
Narayan, R., Paczynski, B., & Piran, T. 1992, ApJL, 395, L83
Nomoto, K. 1982, ApJ, 257, 780
Nomoto, K., & Iben, I., Jr. 1985, ApJ, 297, 531
Oshino, S., Funato, Y., & Makino, J. 2011, PASJ, 63, 881
Pakmor, R., Kromer, M., Taubenberger, S., & Springel, V. 2013, ApJL, 770, L8
Piskunov, A. E., Schilbach, E., Kharchenko, N. V., Röser, S., & Scholz, R. D. 2007, A&A, 468, 151
Plummer, H. C. 1911, MNRAS, 71, 460
Podsiadlowski, P., Langer, N., Poelarends, A. J. T., et al. 2004, ApJ, 612, 1044
Portegies Zwart, S. F., & McMillan, S. L. W. 2000, ApJL, 528, L17
Portegies Zwart, S. F., McMillan, S. L. W., & Gieles, M. 2010, ARA&A, 48, 431
Rastello, S., Amaro-Seoane, P., Arca-Sedda, M., et al. 2019, MNRAS, 483, 1233
Rastello, S., Mapelli, M., Carlo, U. N. D., et al. 2020, MNRAS, 497, 1563
Rodriguez, C. L., Chatterjee, S., & Rasio, F. A. 2016, PhRvD, 93, 084029
Rodriguez, C. L., & Loeb, A. 2018, ApJL, 866, L5
Rosswog, S., Kasen, D., Guillochon, J., & Ramirez-Ruiz, E. 2009, ApJL, 705, L128
Ruiter, A. J., Belczynski, K., Benacquista, M., Larson, S. L., & Williams, G. 2010, ApJ, 717, 1006
Samsing, J., Askar, A., & Giersz, M. 2018, ApJ, 855, 124
Samsing, J., & Ramirez-Ruiz, E. 2017, ApJL, 840, L14
Sana, H., de Mink, S. E., de Koter, A., et al. 2012, Sci, 337, 444
Santoliquido, F., Mapelli, M., Bouffanais, Y., et al. 2020, ApJ, 898, 152
Schwab, J. 2021, ApJ, 906, 53
Shara, M. M., & Hurley, J. R. 2002, ApJ, 571, 830
Shen, K. J. 2015, ApJL, 805, L6
Shen, K. J., & Bildsten, L. 2014, ApJ, 785, 61
Tagawa, H., Kocsis, B., Haiman, Z., et al. 2021, ApJL, 907, L20
Tanikawa, A., Cary, S., Shikauchi, M., Wang, L., & Fujii, M. S. 2024, MNRAS, 527, 4031
Toonen, S., Perets, H. B., Igoshev, A. P., Michaely, E., & Zenati, Y. 2018, A&A, 619, A53
Virtanen, P., Gommers, R., Oliphant, T. E., et al. 2020, NatMe, 17, 261
Wang, J., Hammer, F., & Yang, Y. 2021, MNRAS, 510, 2242
Wang, L., Iwasawa, M., Nitadori, K., & Makino, J. 2020a, MNRAS, 497, 536
Wang, L., Kroupa, P., & Jerabkova, T. 2019, MNRAS, 484, 1843
Wang, L., Nitadori, K., & Makino, J. 2020b, SDAR: Slow-Down Algorithmic Regularization Code for Solving Few-body Problems, Astrophysics Source Code Library, ascl:2002.001
Wang, L., Tanikawa, A., & Fujii, M. 2022, MNRAS, 515, 5106
Webbink, R. F. 1984, ApJ, 277, 355
Willems, B., Kalogera, V., Vecchio, A., et al. 2007, ApJL, 665, L59
Yang, J., Ai, S., Zhang, B.-B., et al. 2022, Natur, 612, 232
Yang, Y.-P., & Zhang, B. 2018, ApJ, 868, 31
Ye, C. S., Fong, W.-f., Kremer, K., et al. 2019, ApJL, 888, L10
Zhang, B. 2020, Natur, 587, 45
Zhong, S.-Q., & Dai, Z.-G. 2020, ApJ, 893, 9